\documentclass[9pt,sigconf,pbalance]{acmart}

\usepackage{algorithm}
\usepackage{algorithmic}
\usepackage{graphicx}
\usepackage{svg}
\usepackage[dvipsnames]{xcolor}
\usepackage{url}
\usepackage{multirow}
\usepackage{tikz}
\usepackage{array}
\usepackage{booktabs}
\usepackage{colortbl}
\usepackage{wrapfig}
\usepackage{siunitx}
\usepackage{listings}
\usepackage{enumitem}
\usepackage[caption=false,font=footnotesize]{subfig}

\usetikzlibrary{positioning, shapes.geometric, arrows.meta, fit, backgrounds, calc}

\usepackage{pifont}
\newcommand{\circnum}[1]{\ding{\the\numexpr181+#1\relax}}

\lstdefinestyle{command}{
  basicstyle=\ttfamily\scriptsize,
  breaklines=true,
  breakatwhitespace=false,
  columns=fullflexible,
  keepspaces=true,
  frame=single
}

\setcopyright{none}
\setcopyright{none}
\makeatletter
\renewcommand\footnotetextcopyrightpermission[1]{}
\makeatother

\acmConference[ASP-DAC'27]{32nd Asia and South Pacific Design Automation Conference}{2027}{Jan}
\acmYear{2027}
\copyrightyear{2027}
\begin{document}

\emergencystretch=1em
\hbadness=10000
\vbadness=10000
\hfuzz=1pt

\title{VPR-Evolve: Multi-Agent-Driven Algorithm Evolution \\ for FPGA Place and Route}

% \author{Qihang Wu}
% \affiliation{%
%   \institution{Arizona State University}
%   \city{Tempe}
%   \state{Arizona}
%   \country{USA}}

% \author{Taizun Jafri}
% \affiliation{%
%   \institution{Arizona State University}
%   \city{Tempe}
%   \state{Arizona}
%   \country{USA}}

% \author{Aman Arora}
% \affiliation{%
%   \institution{Arizona State University}
%   \city{Tempe}
%   \state{Arizona}
%   \country{USA}}

% \author{Vidya A. Chhabria}
% \affiliation{%
%   \institution{Arizona State University}
%   \city{Tempe}
%   \state{Arizona}
%   \country{USA}}

\author{Qihang Wu, Taizun Jafri, Aman Arora, and Vidya A. Chhabria}
\affiliation{%
    \institution{Arizona State University}
    \country{United States}
}

\begin{abstract}
\noindent
	CAD tools typically apply the same fixed, hand-designed algorithms
	across circuits with widely different structural and timing
	characteristics. A common way to specialize these one-size-fits-all flows
	to a target design is to tune the CAD tool’s hyperparameters. However,
	hyperparameter tuning can only select among behaviors already implemented
	by the fixed algorithm, limiting the achievable quality of results while
	requiring many expensive place-and-route evaluations. We present
	VPR-Evolve, a multi-agent framework that specializes Versatile Place and
	Route (VPR), the open-source FPGA pack-place-and-route engine in the
	Verilog-to-Routing (VTR) flow, by evolving its source code for each design.
	VPR-Evolve uses LLM agents to propose, implement, and evaluate code-level
	modifications, while a shared memory records prior outcomes and guides
	subsequent evolution. Every candidate is evaluated through a complete VPR
	build and run, directly optimizing a composite score measured as a weighted
	function of critical-path delay (CPD), routed wirelength (WL), and tool
	runtime (RT). Across five VTR-9 benchmark circuits, VPR-Evolve improves the
	composite score by up to 2.7\% over stock VPR in VTR-9. Relative to stock
	VPR, it reduces CPD by up to 9.8\%, routed WL by up to 18.1\%, and tool RT
	by up to 79.3\%. VPR-Evolve reduces CPD by up to 6.0\%, routed WL by up to
	2.2\%, and tool RT by up to 7.8\% compared with a hyperparameter-tuning baseline.
\end{abstract}

\maketitle

\section{Introduction}
\label{sec:intro}
\noindent
The place-and-route (P\&R) stage of an FPGA CAD flow strongly influences the
quality of results (QoR) of a design, including its critical-path delay (CPD),
routed wirelength (WL), and tool runtime (RT). FPGA designs, however, differ
substantially in their size, connectivity, congestion, and timing
characteristics, while CAD tools typically apply the same hand-designed
algorithms across all designs. These algorithms are therefore broadly
effective compromises rather than specializations to an individual circuit.
The open-source Verilog-to-Routing 9
(VTR~9) flow~\cite{vtr9} and its Versatile Place and Route (VPR)
engine~\cite{betz1997vpr} provide a representative example. VPR transforms a
synthesized FPGA netlist into a placed-and-routed implementation through
packing, placement, and routing. It exposes algorithmic choices and
hyperparameters such as cost-function weights, search schedules, and
move-selection options, but these remain coarse-grained: the underlying
packing, placement, and routing implementations are fixed and applied
uniformly across designs.
Consequently, a single algorithmic structure may not provide the best tradeoff
among CPD, WL, and RT for every circuit.

\noindent\textbf{The limits of hyperparameter tuning.}
A common approach for specializing an FPGA CAD flow to a target design is to
tune the tool's hyperparameters. Frameworks such as
AutoTuner~\cite{autotuner} search the exposed configuration space of an already
implemented algorithm for settings that improve QoR on a particular circuit.
Such a search selects among behaviors the algorithm already supports; it cannot
introduce new mechanisms or modify the source code that defines them, so the
reachable space is bounded by the structure of the fixed algorithm.

Tuning is also evaluation-hungry: each candidate configuration requires
a complete FPGA P\&R run whose cost grows with circuit size, and the many
interacting continuous and categorical parameters that FPGA CAD tools expose
create a space that may need thousands of trials. Tuning therefore specializes
the settings of a fixed tool without jointly exploring the target design and
the algorithms used to implement it.

\noindent\textbf{Key insight.}
FPGAs derive much of their value from their ability to be reconfigured for
different applications. We argue that the CAD algorithms used to configure them
should also be reconfigurable, i.e., the algorithms themselves
can be rewritten and specialized to the target design. In other words, we seek
to reconfigure the algorithms that configure the FPGA. This perspective expands the search from choosing hyperparameters for a fixed
algorithm to exploring code-level algorithm variants. Recent advances in large
language models (LLMs) have made such algorithm evolution possible. LLMs
can reason about source code, propose and implement algorithmic changes, and
revise their proposals from measured outcomes. AlphaEvolve has demonstrated
LLM-driven algorithm evolution in other
domains~\cite{novikov2025alphaevolve}, and~\cite{jafri2026grevolve} introduced
design-adaptive algorithm evolution for ASIC global routing. These works motivate a broader vision of
design--tool co-exploration, in which both the design and the algorithm are variables in the search.

\noindent\textbf{Our approach.}
We present \textbf{VPR-Evolve}, a multi-agent framework that brings
design-adaptive algorithm evolution and design--tool co-exploration~\cite{jafri2026grevolve} to FPGA CAD. VPR-Evolve reconfigures VPR's packing, placement, and routing source code
for each target design. Hyperparameter tuning is retained as an additional
stage within the same loop. VPR-Evolve decomposes the evolution process across multiple LLM agents that
plan, implement, and review candidate changes. A Planner proposes candidate
algorithmic modifications, a Coder implements and benchmarks them, and a
Reviewer decides whether each modification should be retained. An Inspiration
Collector introduces new ideas when the search stops improving. These
agents share a persistent memory of successful and unsuccessful ideas,
previously attempted modifications, measured QoR outcomes, and relevant
algorithmic inspirations, which steers later evolution toward promising
directions and away from repeated unproductive ideas.

The search proceeds through dedicated packing, placement, routing, cross-stage,
and hyperparameter-tuning stages. This staged organization lets
VPR-Evolve reason about individual components of the flow and about
interactions across them. Every candidate is evaluated by rebuilding VPR and performing complete
runs on the target design. The objective is a composite score defined as a
weighted function of CPD, routed WL, and VPR RT. The output is not only
a design-specific configuration but a reconfigured VPR implementation, given as
a reviewable source-code patch. VPR-Evolve also preserves an annotated
evolution trail of which code modifications were attempted and how they
affected QoR. We make the following
contributions:

\begin{itemize}[nosep, leftmargin=*]
	\item To the best of our knowledge, this is the first work to apply
	      LLM-driven evolution to FPGA physical design CAD. VPR-Evolve
	     reconfigures the packing, placement, and routing source code.

	\item We introduce a staged multi-agent framework for algorithm evolution.
	      Planner, Coder, Reviewer, and Inspiration Collector agents cooperate
	      through a shared persistent memory and use measured VPR outcomes to
	      guide algorithm evolution.

	\item We evaluate VPR-Evolve on five VTR~9~\cite{vtr9} benchmark circuits
	      ranging from 37K to 167K primitives. VPR-Evolve improves the
	      composite score, measured as a weighted function of CPD, routed WL,
	      and tool RT, by up to 2.7\% over stock VPR. It reduces CPD by up to
	      9.8\%, routed WL by up to 18.1\%, and tool RT by up to 79.3\%. It
	      reduces CPD by up to 6.0\%, routed WL by up to 2.2\%, and tool RT by
	      up to 7.8\% compared with AutoTuner hyperparameter-tuning baseline
	      with a smaller runtime budget.

	\item We analyze the stage-wise sources of improvement, consistency across
	      random seeds, cross-design transferability, and the reviewability of
	      the source-code modifications.

\end{itemize}

\noindent
VPR-Evolve is fully open-sourced and can be accessed through our GitHub repository~\cite{vpr-evolve-gh}.

\section{Background and Related Work}
\label{sec:background}

\noindent\textbf{FPGA place and route in VPR.}
VPR~\cite{betz1997vpr}, the pack, place, and route engine of
VTR~9~\cite{vtr9}, transforms a synthesized netlist into a placed-and-routed implementation on a target
FPGA architecture. Packing is performed by AAPack~\cite{luu2011aapack}, which greedily groups LUTs, 
flip-flops, memories, and other primitives into legal clustered logic blocks. Placement uses simulated
annealing with a cost balancing estimated wirelength and delay. VTR~9 additionally uses a
reinforcement-learning agent to select among available placement move types
and provides an experimental analytical placement flow that jointly considers
primitive-level packing and placement. Routing uses a timing-driven
negotiated-congestion algorithm based on PathFinder~\cite{mcmurchie1995pathfinder} which 
routes nets using directed shortest-path search while balancing congestion and
path delay. Although VPR exposes algorithmic choices
and hyperparameters within these stages, their underlying implementations
remain fixed once selected and are applied uniformly across designs.
VPR-Evolve instead modifies these implementations per design, letting
the resulting flow combine mechanisms drawn from multiple approaches in the
FPGA CAD literature rather than relying on one predefined algorithm.

\noindent\textbf{Hyperparameter tuning for CAD.}
A common approach for adapting a fixed EDA flow to a target design is to tune
the tool's exposed hyperparameters. Prior methods have used Bayesian
optimization~\cite{autotuner,optuna}, reinforcement learning~\cite{agnesina2020rl}, and parallel multi-armed-bandit search, as in
DATuner~\cite{xu2017datuner}, to identify parameter configurations that
improve CPD. These methods differ in how they explore the parameter
space, but all remain constrained to the behaviors exposed by an existing CAD
implementation. We use AutoTuner as a representative hyperparameter-tuning
baseline and instantiate it with Optuna's TPE optimizer. Our primary
comparison is between tuning a fixed implementation and modifying its source
code, so the specific tuning strategy is secondary to the search space it can
access. VPR-Evolve expands this space by evolving the algorithms themselves,
and tuning remains complementary: it can be applied afterward to optimize the
exposed parameters of the evolved implementation.

\noindent\textbf{Algorithm evolution for EDA.}
A recent line of work removes the fixed-implementation constraint by
letting LLM agents modify tool source code directly.
SATLUTION~\cite{yu2025satlution} scales LLM-based code evolution to full
repository scope, evolving SAT solver codebases under strict correctness
guarantees while also evolving its own evolution policies. Within EDA,
AuDoPEDA~\cite{ghose2026audopeda} deploys an autonomous coding agent that
reads the OpenROAD codebase, proposes research directions, and submits
executable diffs in a closed loop driven by QoR, and ABC/Yosys
Evolve~\cite{yu2026abcevolve} brings the approach to logic synthesis, with
multiple LLM agents iteratively rewriting the integrated ABC codebase. 
GR-Evolve~\cite{jafri2026grevolve} extends this paradigm to EDA by applying
LLM-driven algorithm evolution to global routing in the OpenROAD flow.
VPR-Evolve introduces this evolution to FPGA CAD, where the flow spans
multiple interacting stages (packing, placement, and routing) and every
candidate must be rebuilt and evaluated through the complete flow. It evolves
VPR for each target design and folds hyperparameter tuning into the same loop
as an additional stage.

\section{VPR-Evolve Overview}
\label{sec:overview}
\begin{figure}[t]
	\centering
	\includegraphics[width=\columnwidth]{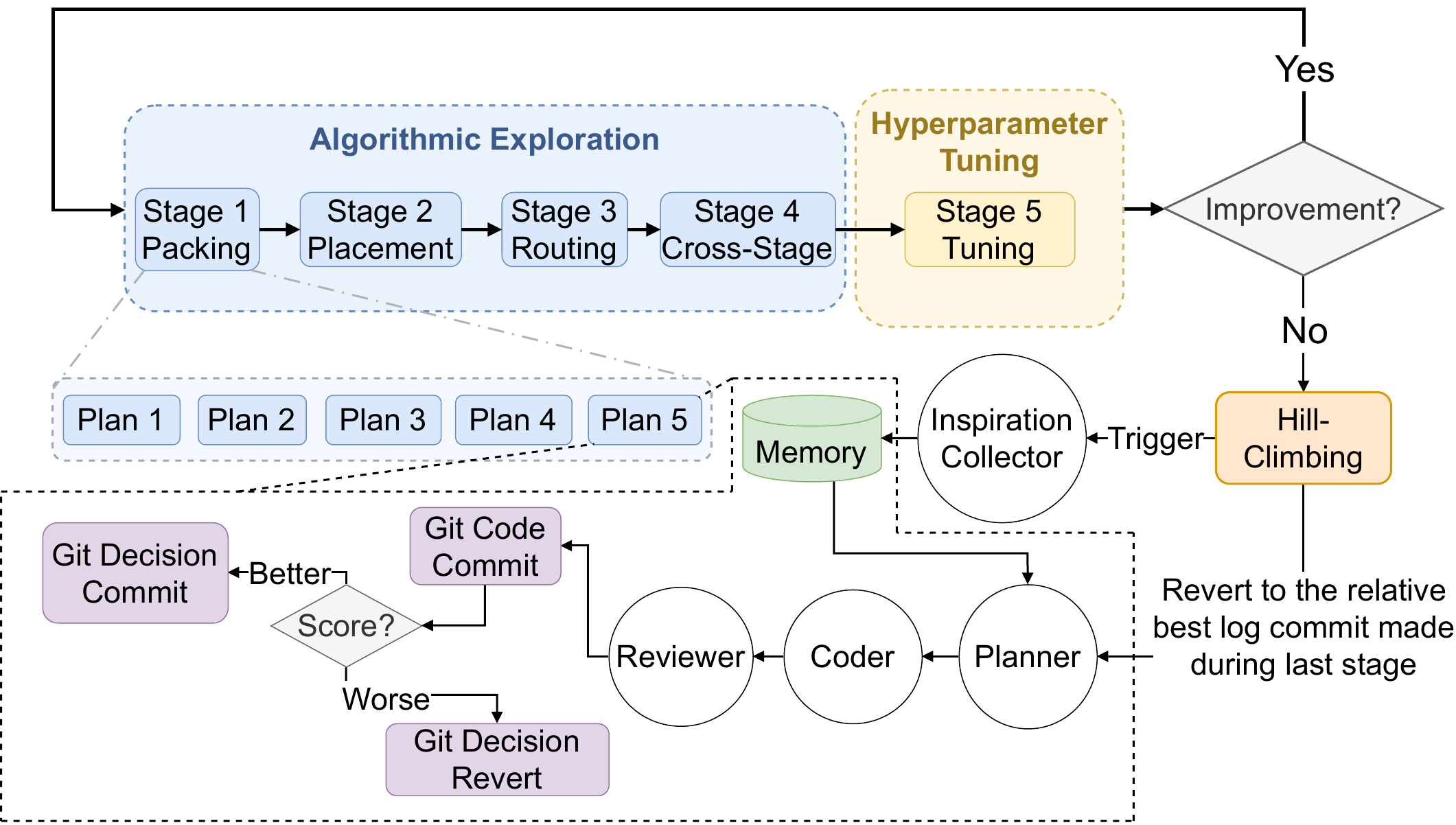}
    \vspace{-7mm}
	\caption{Overview of VPR-Evolve Loop.}
    \vspace{-12mm}
	\label{fig:overview}
\end{figure}

\noindent
VPR-Evolve specializes VPR to a target FPGA design through a closed loop of
proposing, implementing, and evaluating source-code modifications. As shown in
Figure~\ref{fig:overview}, the search begins from stock VPR and traverses five
stages in each iteration loop: packing, placement, routing, cross-stage evolution, and
hyperparameter tuning. Each candidate is evaluated using complete VPR runs on
the target design and scored using CPD, WL, and tool runtime.

A \emph{Planner} agent proposes code modifications, a \emph{Coder} agent implements and evaluates them, and a
\emph{Reviewer} agent records whether each candidate is retained or reverted. The agents
share a persistent memory of prior ideas and measured outcomes. When progress
stalls, an \emph{Inspiration Collector} adds new algorithmic ideas and the search may
restart from a promising reverted candidate. The final output is the best
design-specialized VPR implementation, its hyperparameter configuration, and
an annotated git history.

\section{VPR-Evolve Framework Details}
\label{sec:methodology}
\noindent
Algorithm~\ref{alg:main} formalizes the complete evolution process. The search
starts from stock VPR and traverses the ordered stage sequence $ S=\langle
\textsc{Pack}, \textsc{Place}, \textsc{Route}, \textsc{CrossStage},
\textsc{Tune} \rangle . $ For each stage, the Planner generates up to $N$
modification plans, and the Coder evaluates each plan using $B$ in-loop seeds.
Before evolution begins, VPR-Evolve compares the simulated-annealing (SA) and
analytical-placement (AP) flows using $C$ seeds and selects the better flow for
the placement stage. The search is bounded by an iteration budget $I_{\max}$,
wall-time budget $T_{\max}$, plateau-escape threshold $P_{\mathrm{escape}}$,
and maximum number of plateau escapes $E_{\max}$. LLM usage is additionally
constrained by hourly and weekly token budgets. Reaching the hourly limit
pauses execution until the window resets, whereas reaching the weekly limit
terminates the search. The variable $x$ denotes the current search state from
which candidates are generated, while $x^{*}$ denotes the best implementation
encountered. 

\begin{algorithm}[!t]
	\small
	\caption{VPR-Evolve evolution loop.}
	\label{alg:main}
    % \makeatletter
    % \let\saved@fs@post\@fs@post
    % \def\@fs@post{\saved@fs@post\kern-5em}
    % \makeatother
	\begin{algorithmic}[1]

		\REQUIRE VPR codebase $VPR$, memory $\mathcal{M}$, stages
		$S=\langle\textsc{Pack},\textsc{Place},\textsc{Route},
			\textsc{CrossStage},\textsc{Tune}\rangle$
		\REQUIRE plans per stage $N$, seed counts $B$ and $C$, budgets
		$I_{\max}$, $E_{\max}$, $P_{\mathrm{escape}}$, and $T_{\max}$

		\STATE $flow \gets
			\textsc{ComparePlacementFlows}(\{\textsc{SA},\textsc{AP}\},C)$
		\STATE $x \gets VPR$; $x^{*}\gets VPR$
		\COMMENT{current and best implementations}
		\STATE $itr\gets0$; $plateau\gets0$; $escapes\gets0$
		\STATE $r^{*}\gets\varnothing$
		\COMMENT{best valid candidate rejected during plateau}

		\WHILE{$itr<I_{\max}$ \AND $escapes<E_{\max}$
			\AND $\textsc{Elapsed}()<T_{\max}$
			\AND $\textsc{TokenBudgetAvailable}()$}

		\STATE $kept\gets0$

		\FOR{each $s\in S$}
		\STATE $\textsc{WaitIfHourlyLimitReached}()$
		\STATE $P\gets\textsc{Planner}(s,\mathcal{M},x,N)$

		\FOR{each $p\in P$}
		\STATE $c\gets\textsc{Coder}(p,s,flow,x,B)$

		\IF{$\textsc{Valid}(c)$ \AND
			$\textsc{Score}(c)<\textsc{Score}(x)$}

		\STATE $x\gets\textsc{Keep}(c)$; $kept\gets kept+1$

		\IF{$\textsc{Score}(x)<\textsc{Score}(x^{*})$}
		\STATE $x^{*}\gets x$
		\ENDIF

		\ELSE
		\IF{$\textsc{Valid}(c)$}
		\STATE $r^{*}\gets
			\textsc{BetterOf}(r^{*},c)$
		\ENDIF
		\STATE $\textsc{RevertAndRecord}(c)$
		\ENDIF
		\ENDFOR
		\ENDFOR

		\IF{$kept>0$}
		\STATE $plateau\gets0$
		\STATE $r^{*}\gets\varnothing$
		\COMMENT{start a new plateau period}

		\ELSE
		\STATE $plateau\gets plateau+1$

		\IF{$plateau\geq P_{\mathrm{escape}}$
			\AND $r^{*}\neq\varnothing$}

		\STATE $x\gets\textsc{PlateauEscape}(r^{*})$
		\COMMENT{controlled non-improving move}
		\STATE $\textsc{InspirationCollector}(\mathcal{M})$
		\STATE $plateau\gets0$; $escapes\gets escapes+1$; $r^{*}\gets\varnothing$
		\ENDIF
		\ENDIF

		\STATE $itr\gets itr+1$

		\ENDWHILE

		\RETURN $x^{*}$ and the annotated git history

	\end{algorithmic}
% \par\vspace{-2em}
\end{algorithm}

\subsection{Multi-agent Architecture}
\label{sec:meth:agents}
\noindent
VPR-Evolve separates proposal, implementation, and evaluation across three
primary LLM agents: the Planner, Coder, and Reviewer. Each agent
receives only the context and permissions its role requires, which limits
per-agent context and prevents any agent from judging its own modifications. 
A fourth agent, the Inspiration
Collector, is invoked only during plateau recovery.

\noindent\textbf{Planner.}
The Planner proposes code-level modifications for the current evolution stage.
It receives the stage-relevant VPR source, the current search state $x$, the
stage-specific idea menu, prior outcomes stored in the shared memory
$\mathcal{M}$, and an algorithmic source from the inspiration library. For each
plan, it identifies the mechanism to modify, the expected QoR impact, the
relevant source-code region, implementation feasibility, and potential risks.
It also checks whether the same or a closely related idea has already
been attempted, and produces at most $N$ plans per stage.

\noindent\textbf{Coder.}
The Coder receives a single plan and implements it within an isolated VPR worktree. It
edits the source code permitted by the current stage, rebuilds VPR, and performs
complete runs using the $B$ in-loop seeds.
During the placement stage, the Coder uses the placement flow selected by
\textsc{ComparePlacementFlows}. It returns the source-code diff, build and
execution status, composite score, and the corresponding CPD, routed WL, and RT.

\noindent\textbf{Reviewer.}
The Reviewer receives the candidate diff, build and execution status, measured
QoR, and the current search state. A candidate that fails to compile or
complete P\&R is labeled \texttt{BUILD\_FAILED}. A valid candidate is labeled
\texttt{KEPT} when it improves the composite score of the current state;
otherwise, it is labeled \texttt{REVERTED}.
% The Reviewer decides; it does not write code.

\noindent\textbf{Inspiration Collector.}
It is activated only when the search invokes plateau
recovery. It searches relevant FPGA CAD literature and open-source reference
implementations, summarizes potentially useful mechanisms, and appends new
ideas to the inspiration library and stage-specific idea menus. It may update
the shared memory, but it cannot modify source code or make acceptance
decisions.

\subsection{Cost Function Scoring}
\label{sec:meth:score}
\noindent
Every candidate is judged by a score computed from a weighted cost
function of CPD, WL, and VPR runtime. For each QoR metric, let $S_b$ denote the baseline
value obtained from the fixed VPR implementation and $S_m$ denote the value
measured for a candidate implementation. The relative improvement,
expressed in percent, is $\mathrm{imp}(S_b,S_m) = 100 \cdot
\frac{S_b-S_m}{S_b}$. The composite score is $\mathrm{Score} = S_{\mathrm{ref}}
- \left( \alpha\,\mathrm{imp}_{\mathrm{CPD}} +
\beta\,\mathrm{imp}_{\mathrm{WL}} + \gamma\,\mathrm{imp}_{\mathrm{RT}}
\right)$. The weights $\alpha$, $\beta$, and $\gamma$ control the relative
importance of CPD, WL, and RT, respectively, and satisfy
$\alpha+\beta+\gamma=1$. We include runtime in the score to ensure that CPD and WL are not improved at the cost of runtime increases. The baseline values $S_b$ are
measured once per circuit from stock VPR at search setup and remain fixed for
the rest of the run; the recorded per-circuit reference values are included in
the released repository. The
candidate measurements and selected weights are defined in
Section~\ref{sec:exp}. The fixed VPR implementation has zero relative
improvement for all three metrics and therefore receives a composite score of
exactly $S_{\mathrm{ref}}$. 

\subsection{Staged Algorithm Evolution}
\label{sec:meth:search}
\noindent
Each iteration traverses five stages in a fixed order. Stage S1 evolves the
packing implementation, S2 evolves placement, and S3 evolves routing. Changes
within these stages are restricted to the source code associated with the
corresponding VPR component. Stage S4 performs cross-stage evolution, during
which a plan may modify any combination of packing, placement, and routing to
capture interactions across stage boundaries. S4 also includes plans that
port a published algorithm from the inspiration library into VPR.
Stage S5 tunes the exposed
hyperparameters of the implementation produced by the preceding algorithmic
stages. For the placement stage, evolution is performed only on the placement
flow selected before the main loop. Within each stage, plans are evaluated
sequentially. A valid candidate replaces the current search state $x$ only when
it achieves a lower composite score. Consequently, an accepted change becomes
the starting point for all subsequent plans. Rejected candidates are restored,
while candidates that fail to build or complete P\&R are marked as invalid. The
best-so-far implementation $x^{*}$ is maintained separately from the current
search state. Whenever an accepted candidate improves upon $x^{*}$, it becomes
the new best-so-far implementation. Each evaluated plan is recorded in an
annotated git history together with its source-code diff, measured QoR, and
outcome: \texttt{KEPT}, \texttt{REVERTED}, or \texttt{BUILD\_FAILED}. This
history provides the evolution trail analyzed in Section~\ref{sec:discussion}.

\subsection{Shared Memory}
\label{sec:meth:memory}
\noindent
VPR-Evolve maintains a persistent shared memory $\mathcal{M}$ across agents,
stages, and iterations. Its long-term portion contains the inspiration library,
summaries of relevant FPGA CAD literature and open-source reference
implementations, stage-specific idea menus, and the annotated history of
previously evaluated plans.

Its short-term portion records the current search state and score, the
best-so-far score and CPD, recently evaluated plans and their outcomes, and the
idea tags already attempted during the current iteration. The Planner uses this
information to generate proposals that are relevant to the current search state
while avoiding repeated exploration of previously unsuccessful ideas.

\subsection{Plateau Recovery and Termination}
\label{sec:meth:termination}
\noindent
An iteration in which no candidate is retained is classified as a plateau
iteration. The counter $p$ records the number of consecutive plateau iterations
and is reset whenever at least one candidate is accepted. During this period,
VPR-Evolve tracks the best valid reverted candidate $r^{*}$ encountered since
the last accepted iteration or plateau escape. When $p$ reaches the
plateau-escape threshold $P_{\mathrm{escape}}$, VPR-Evolve performs a
controlled non-monotonic move. The candidate $r^{*}$ becomes the new current
search state $x$, even though it did not improve upon the previous state. This
move allows subsequent plans to explore from a different implementation without
replacing the best-so-far solution $x^{*}$.

The Inspiration Collector is invoked during the same recovery operation. It
gathers additional mechanisms from FPGA CAD literature and reference
implementations and appends them to the shared memory. The plateau counter is
then reset, the escape counter is incremented, and the next iteration proceeds
from the recovered state using the refreshed idea pool. The search terminates
when the iteration count reaches $I_{\max}$, the number of plateau escapes
reaches $E_{\max}$, the elapsed wall time reaches $T_{\max}$, or the weekly
token budget is exhausted. The hourly token limit is treated as a rate limit:
execution pauses until the hourly window resets and then resumes without
advancing the iteration counter. At termination, VPR-Evolve returns the
best-so-far implementation $x^{*}$ and its annotated evolution history.
% \subsection{Design choices}
% \label{sec:meth:design}

% \noindent\textbf{Why specialize per design.}
% A one-size-fits-all heuristic is a compromise across a benchmark suite, not the
% right choice for any single circuit (Section~\ref{sec:intro}). VPR-Evolve
% therefore evolves against one target design at a time, so the score it optimizes
% reflects that circuit alone. Whether the resulting recipes carry over to other
% circuits is a separate question, examined in
% Section~\ref{sec:results:transfer}.

% \noindent\textbf{Why evaluate in the loop on real runs.}
% Every candidate is scored by a real VPR build and a full P\&R on the target design
% rather than by a surrogate model. A surrogate would be cheaper per candidate but
% would have to be trusted precisely where the search pushes hardest, on algorithm
% variants no surrogate was trained on. Measuring real QoR keeps the search honest
% at the cost of a rebuild per plan, a cost the loop offsets with sample efficiency
% (Section~\ref{sec:results:time}).
%
% \noindent\textbf{Why a single monotonic incumbent.}
% Rather than maintain a population as a genetic algorithm would, VPR-Evolve keeps a
% single incumbent and either commits or reverts each plan against it, preserving
% rejects in git and hill-climbing off plateaus. This keeps the output of a run a
% single, reviewable patch series rather than a population of variants, and it lets
% a designer read, accept, or re-run any individual change
% (Section~\ref{sec:discussion:trail}).

\section{Experimental Setup}
\label{sec:exp}

\noindent
{\bf Benchmarks.} We evaluate on five large benchmark circuits from the VTR~9 benchmark
suite~\cite{vtr9} (Table~\ref{tab:benchmarks}). VPR-Evolve shines on
larger benchmarks, where hyperparameter tuning struggles; on small designs the
scope for improvement is narrow and tuning can yield similar benefits.

\begin{table}[t]
	\centering
	\caption{Five VTR~9 benchmark circuits used for evaluation.}
    \vspace{-4mm}
	\label{tab:benchmarks}
	\small
	\begin{tabular}{l r r r c r}
		\toprule
		Circuit & LUTs   & FFs    & Primitives & Layout         & $W$ \\
		\midrule
		LU8     & 24,632 & 6,224  & 36,833     & $68\times68$   & 108 \\
		bgm     & 30,231 & 5,105  & 37,617     & $76\times76$   & 88  \\
		sv2     & 25,552 & 15,585 & 53,686     & $96\times96$   & 116 \\
		LU32    & 84,884 & 19,668 & 123,305    & $123\times123$ & 148 \\
		mcml    & 90,208 & 51,529 & 167,249    & $117\times117$ & 164 \\
		\bottomrule
	\end{tabular}
     \vspace{-4mm}
\end{table}

\noindent
{\bf VPR and VTR settings.} We use VPR's flagship architecture
\\ \texttt{k6\_frac\_N10\_frac\_chain\_mem32K\_40nm} with a fixed layout and a fixed channel width, which are determined once per circuit from VPR's auto minimum layout and channel width searching: the layout dimension is
$1.2 \times d_{\min}$ for the minimum auto-layout dimension $d_{\min}$, and the routing channel width is $1.2 \times W_{\min}$ rounded up to the next even value, since channel width only takes even numbers.
This results in approx. 80\% resource usage, ensuring a challenging problem for the CAD tool and mirroring common industrial FPGA deployment scenarios.
Every experiment runs on this same fixed layout, so no gains are from device sizing.

\noindent
{\bf Hardware evaluation platform.} Our experiments are performed on a datacenter server with a 64-core CPU and 256\,GB of RAM.

\noindent
{\bf LLM configuration.} All agents run on Anthropic Claude models through
the Claude Code agent harness. The Planner runs on Claude Opus~4.7 with a 1M-token
context window, since plan generation is the one judgment-heavy call per stage; the
Coder, the Reviewer, and the Inspiration Collector run on Claude Sonnet~4.6.

\noindent
{\bf Baselines.} We compare against two baselines. The first is default stock VTR~9 pinned commit   \texttt{bd5434a94}, so all evolved algorithms and tuned
configurations are measured on an identical code and baseline. The second baseline is the AutoTuner hyperparameter-tuning framework~\cite{autotuner}. We run AutoTuner, driven by Ray Tune~\cite{liaw2018tune}, with
Optuna's Tree-structured Parzen Estimator (TPE) sampler~\cite{optuna,
bergstra2011tpe}, a Bayesian-family sequential model-based optimizer. Both baselines are evaluated with 100 seeds at the end of tuning or evolution. Each iteration during training and evolution uses $B=5$ seeds for QoR. All composite scores, both in-loop and in Section~\ref{sec:results}, are computed against the same fixed per-circuit stock baseline recorded at search setup. Reported
scores are means of the per-seed scores, while Table~\ref{tab:qor} reports
geometric means of CPD, WL, and RT. The parameters used in VPR-Evolve are in Table~\ref{tab:parameters}.
%We choose $\alpha=0.5$, $\beta=0.2$ and $\gamma=0.3$ to 

\begin{table}[t]
\centering
\caption{VPR-Evolve parameters.}
\vspace{-4mm}
\label{tab:parameters}
\scriptsize
\setlength{\tabcolsep}{3pt}
\renewcommand{\arraystretch}{1.05}
\begin{tabular}{p{0.20\columnwidth}p{0.54\columnwidth}p{0.16\columnwidth}}
\toprule
\textbf{Parameter} & \textbf{Definition} & \textbf{Value} \\
\midrule
$N$ & Maximum plans proposed per stage & 5 \\
$B$ & In-loop seeds used to evaluate each candidate & 5 \\
$C$ & Seeds used to compare SA and AP placement & 100 \\
$I_{\max}$ & Maximum number of evolution iterations & 5 \\
$P_{\mathrm{escape}}$ & Plateau iterations before recovery & 1 \\
$E_{\max}$ & Maximum number of plateau escapes & 2 \\
$T_{\max}$ & Wall-time budget per circuit & 120 hrs \\
$Token_{\mathrm{hourly}}$ & Hourly token limit that pauses the search & 60M \\
$Token_{\mathrm{weekly}}$ & Weekly token budget that terminates the search & 600M \\
\midrule
$\alpha$ & CPD weight in the composite score & 0.5 \\
$\beta$ & Wirelength weight in the composite score & 0.2 \\
$\gamma$ & Runtime weight in the composite score & 0.3 \\
$S_{\mathrm{ref}}$ & Reference score assigned to fixed VPR & 1000 \\
\bottomrule
\end{tabular}
\vspace{-4mm}
\end{table}

\section{VPR-Evolve Evaluation}
\label{sec:results}

\subsection{QoR Against AutoTuner and VTR~9}
\label{sec:results:qor}
\noindent
We give AutoTuner a generous budget of up to 3,000 iterations per
circuit, which costs more tuning wall-clock than VPR-Evolve on every circuit.

\begin{figure}[t]
	\centering
	\includegraphics[width=\columnwidth]{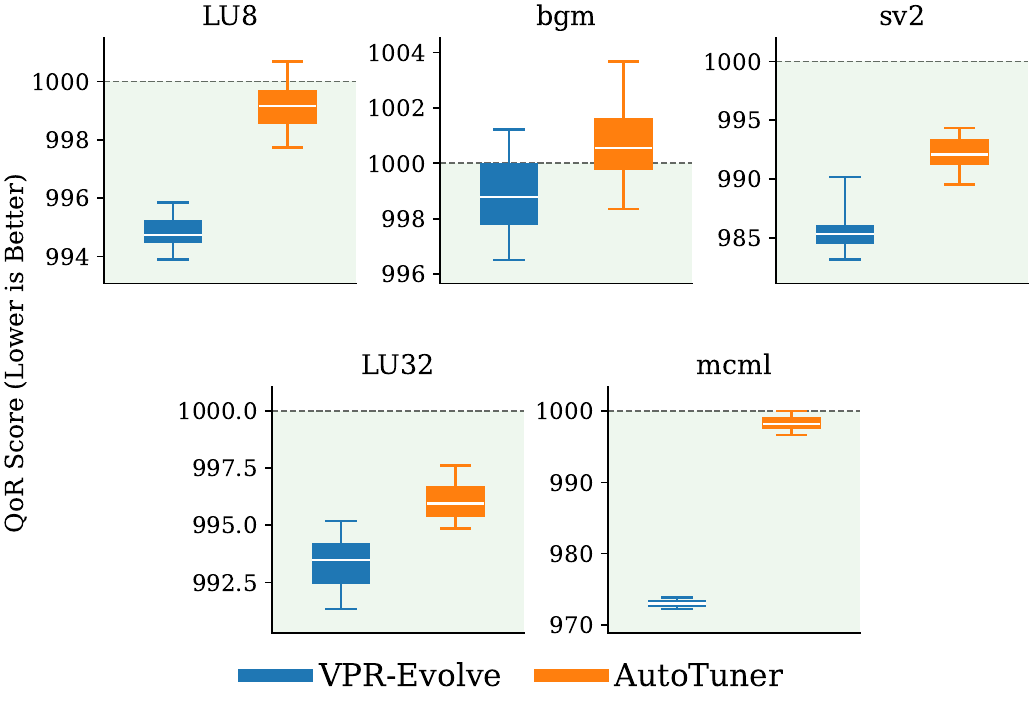}
    \vspace{-8mm}
	\caption{Composite score over the 100 evaluation seeds, one panel per
		circuit (1000 = default VTR
		9 reference).}
        \vspace{-10mm}
	\label{fig:qor-score}
\end{figure}

\noindent\textbf{Composite QoR.}
Figure~\ref{fig:qor-score} shows the composite-score distributions over the
100 independent evaluation seeds. Each box spans the 25th--75th percentiles,
the center line denotes the median, and the whiskers denote the 5th--95th
percentiles. The dashed line at 1000 represents the default VTR-9 reference;
lower values indicate better QoR. VPR-Evolve's score distribution lies below
AutoTuner and the default VTR-9 reference on every circuit.  The boxes and
whiskers further show that these improvements are
consistent across independent seeds. The coefficient of variation (CoV) of
VPR-Evolve's composite score
is 0.06\%, 0.14\%, 0.20\%, 0.05\%, and 0.12\% on \texttt{LU8},
\texttt{bgm}, \texttt{sv2}, \texttt{mcml}, and \texttt{LU32},
respectively.

\begin{table}[t]
    \centering
    \caption{Per-circuit CPD, WL, and tool runtime, reported as geometric means across 100
    evaluation seeds. The default RT column is the fixed per-circuit stock
    baseline.}
    \vspace{-5mm}
    \label{tab:qor}
    \small
    \resizebox{\columnwidth}{!}{%
        \begin{tabular}{l ccc c ccc ccc}
            \toprule
            & \multicolumn{3}{c}{Default VTR-9 VPR }
            &
            & \multicolumn{3}{c}{VPR-Evolve}
            & \multicolumn{3}{c}{AutoTuner} \\
            \cmidrule(lr){2-4}
            \cmidrule(lr){6-8}
            \cmidrule(lr){9-11}

            Circuit
            & \shortstack{CPD\\(ns)}
            & WL
            & \shortstack{RT\\(s)}
            & \shortstack{Placer \\ Winner }
            & \shortstack{CPD\\(ns)}
            & WL
            & \shortstack{RT\\(s)}
            & \shortstack{CPD\\(ns)}
            & WL
            & \shortstack{RT\\(s)} \\
            \midrule

            LU8
            & 73.55
            & \textbf{\color{ForestGreen}299,621}
            & 101
            & SA
            & \textbf{\color{ForestGreen}73.24}
            & 306,777
            & \textbf{\color{ForestGreen}83}
            & 74.48
            & 301,392
            & 90 \\

            bgm
            & 20.20
            & 358,086
            & 106
            & SA
            & \textbf{\color{ForestGreen}19.96}
            & 355,893
            & \textbf{\color{ForestGreen}103}
            & 21.13
            & \textbf{\color{ForestGreen}328,997}
            & 110 \\

            sv2
            & 14.47
            & 372,893
            & 98
            & AP
            & \textbf{\color{ForestGreen}13.05}
            & 305,466
            & 80
            & 13.34
            & \textbf{\color{ForestGreen}301,609}
            & \textbf{\color{ForestGreen}69} \\

            LU32
            & 75.66
            & 1,219,834
            & 521
            & AP
            & \textbf{\color{ForestGreen}75.03}
            & \textbf{\color{ForestGreen}1,125,196}
            & 439
            & 75.49
            & 1,150,584
            & \textbf{\color{ForestGreen}394} \\

            mcml
            & 48.58
            & 910,450
            & 2,175$^{\dagger}$
            & AP
            & \textbf{\color{ForestGreen}46.81}
            & \textbf{\color{ForestGreen}834,429}
            & \textbf{\color{ForestGreen}451}
            & 49.78
            & 841,805
            & 461 \\

            \bottomrule
        \end{tabular}%
  }

    \vspace{2pt}
    \parbox{\columnwidth}{\footnotesize $^{\dagger}$ Dominated by
    timed-out runs: on most seeds stock VPR fails to route \texttt{mcml} at our
    fixed layout and channel width and stops at the 3000\,s timeout.}
    \vspace{-5mm}
\end{table}

\noindent\textbf{Individual QoR metrics.}
Table~\ref{tab:qor} decomposes the composite score into CPD, WL, and tool RT. VPR-Evolve
achieves the lowest geometric-mean CPD on every circuit.
%Relative to
% AutoTuner, it reduces CPD by 1.7\% on \texttt{LU8}, 5.5\% on \texttt{bgm},
% 2.2\% on \texttt{sv2}, and 6.0\% on \texttt{mcml}. Relative to default
% VTR~9, its largest CPD reduction is 9.8\% on \texttt{sv2}.
The WL and RT columns expose the tradeoffs selected by the composite objective.
AutoTuner obtains lower WL on \texttt{LU8}, \texttt{bgm}, and \texttt{sv2},
whereas VPR-Evolve obtains 2.2\% lower WL on \texttt{LU32} and 0.9\% lower WL on
\texttt{mcml}. Nevertheless,
VPR-Evolve's stronger CPD and RT results produce a lower composite score on all
circuits. Compared with AutoTuner, VPR-Evolve reduces per-seed VPR RT by 7.8\%
on \texttt{LU8}, 6.4\% on \texttt{bgm}, and 2.2\% on \texttt{mcml}. On
\texttt{sv2} and \texttt{LU32} it accepts a higher RT (15.9\% and 11.4\%,
respectively) in exchange for lower CPD, and on both circuits the composite
score still favors VPR-Evolve. Compared with default VTR~9, VPR-Evolve reduces routed WL by up to
18.1\% and per-seed VPR RT by up to 79.3\%. These results show that VPR-Evolve
reaches QoR tradeoffs that are not obtained by tuning the exposed parameters of
the fixed algorithms in VPR.

\begin{figure}[t]
	\centering
	\includegraphics[width=0.68\columnwidth]{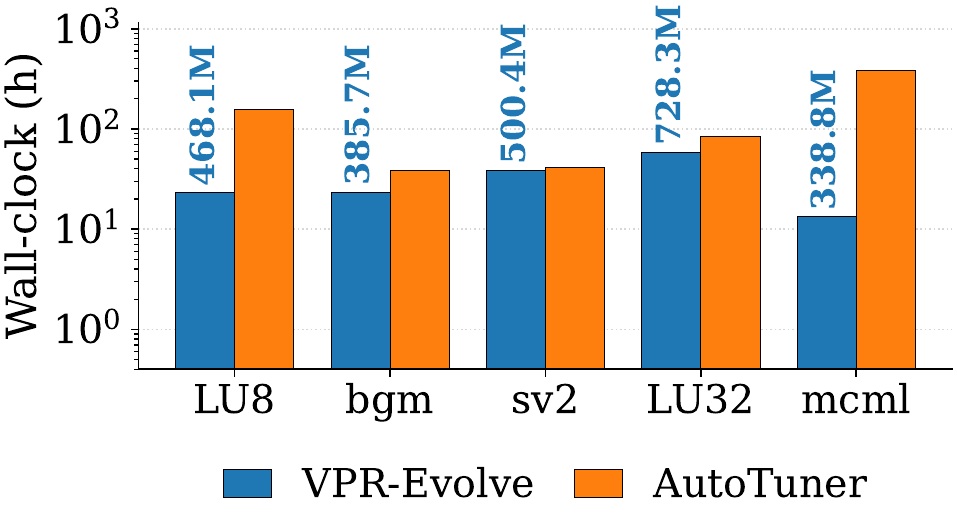}
    \vspace{-7mm}
	\caption{Total tuning wall-clock of VPR-Evolve. Blue
		label above each VPR-Evolve bar is its total LLM token count used.}
        \vspace{-12mm}
	\label{fig:walltime}
\end{figure}

\noindent\textbf{Parameter tuning and evolution runtime.}
Figure~\ref{fig:walltime} reports the end-to-end tuning wall-clock required to
produce the final evolved implementation and the 3000-iteration AutoTuner result. Unlike the per-seed VPR RT in Table~\ref{tab:qor}, this metric captures the
total cost of overall flow. VPR-Evolve requires less tuning wall-clock than AutoTuner on every circuit while achieving better composite QoR (Table~\ref{tab:qor}). This advantage primarily
results from search efficiency. For example, on \texttt{mcml}, VPR-Evolve evaluates only
25 code modifications, compared with 2,645 AutoTuner configurations. Across
five in-loop seeds, this corresponds to 125 versus 13,225 VPR runs, or
approximately 106$\times$ fewer evaluations.  Although algorithm evolution incurs a substantial one-time search cost, the
resulting implementation can be reused across many subsequent VPR runs.
Conceptually, VPR-Evolve produces a new release of the place-and-route tool
specialized to the target design. Unlike a conventional release, which
provides one implementation for many designs, this evolved release contains
algorithms customized to a single design. The tuning wall-clock in
Fig.~\ref{fig:walltime} should therefore be interpreted as a one-time
design-specific tool-development cost, whereas the per-seed VPR runtime in
Table~\ref{tab:qor} represents the cost of repeatedly deploying the resulting
tool. The labels in Fig.~\ref{fig:walltime} report the LLM
tokens consumed by each VPR-Evolve run. These totals are cumulative
over the whole run and are dominated by cached-context reads, which the
provider serves at reduced cost and does not count to token budgets
in Table~\ref{tab:parameters}. Input, output, and cache-creation tokens
stay below $Token_{\mathrm{weekly}}$, so every run ended on $I_{\max}$,
$E_{\max}$, or $T_{\max}$ rather than the weekly limit.

% The evolved executable can be reused across seeds, repeated runs, and design
% revisions, allowing this one-time design-specific tool-development cost to be
% amortized. The figure also reports LLM token usage, which depends more on the
% number of explored algorithmic directions than on circuit size. 

% This advantage arises from search efficiency rather than faster individual
% P\&R runs. On \texttt{mcml}, VPR-Evolve evaluates 25 candidate code
% modifications, whereas AutoTuner evaluates 2,645 candidate hyperparameter
% configurations. Because both methods evaluate every candidate over the same
% five in-loop seeds, these searches correspond to 125 and 13,225 complete P\&R
% runs, respectively. VPR-Evolve therefore uses approximately 106$\times$ fewer
% candidate evaluations and P\&R runs while obtaining a lower composite score.

\subsection{Cross-Design Transferability}
\label{sec:results:transfer}
\noindent
We evaluate whether implementations evolved for one source circuit transfer to
other circuits without additional evolution. Table~\ref{tab:transfer-score}
reports the composite score obtained by applying each evolved implementation
unchanged to every target circuit. Rows denote the source circuit and columns
the target circuit. Scores are normalized to default VTR-9, with values below
1000, indicating improvement. Diagonal entries represent native evaluation,
while off-diagonal entries measure cross-design transfer.

Among the 19 completed off-diagonal transfers, 17 improve over default VTR-9
VPR. The implementations evolved on \texttt{LU8}, \texttt{sv2}, and
\texttt{mcml} improve all completed target circuits, showing that
VPR-Evolve discovers code-level improvements
that are not limited to the source design. For \texttt{sv2} and \texttt{mcml}
the largest gains occur on the source design itself, while for \texttt{LU8},
\texttt{bgm}, and \texttt{LU32} at least one transferred implementation
matches or exceeds the source evolved result. These results establish that VPR-Evolve can produce both
design-specialized and broadly reusable algorithmic improvements.

\begin{table}[t]
	\centering
	\caption{Cross-circuit transfer composite score (lower is better; 1000 = VTR~9).
		Rows are the \emph{source} circuit whose evolved recipe is applied to the \emph{target}
		column.}
        \vspace{-4mm}
	\label{tab:transfer-score}
	\small
	\begin{tabular}{l ccccc}
		\toprule
		               & \multicolumn{5}{c}{\textbf{Transferring Design}}                                                                                                                             \\
		\cmidrule(lr){2-6}
		Evolved Source & LU8                                   & bgm                                   & sv2                                   & LU32         & mcml                                  \\
		\midrule
		LU8            & \textbf{\textcolor{ForestGreen}{995}} & \textcolor{ForestGreen}{994}          & \textcolor{ForestGreen}{995}          & \textcolor{ForestGreen}{990}          & \textcolor{ForestGreen}{990}          \\
		bgm            & \textcolor{ForestGreen}{992}          & \textbf{\textcolor{ForestGreen}{999}} & 1011                                  & \textcolor{ForestGreen}{993}          & --                                    \\
		sv2            & \textcolor{ForestGreen}{991}          & \textcolor{ForestGreen}{997}          & \textbf{\textcolor{ForestGreen}{986}} & \textcolor{ForestGreen}{987}          & \textcolor{ForestGreen}{991}          \\
		LU32           & \textcolor{ForestGreen}{993}          & 1003                                  & \textcolor{ForestGreen}{993}          & \textbf{\textcolor{ForestGreen}{993}} & \textcolor{ForestGreen}{992}          \\
		mcml           & \textcolor{ForestGreen}{993}          & \textcolor{ForestGreen}{995}          & \textcolor{ForestGreen}{994}          & \textcolor{ForestGreen}{988}          & \textbf{\textcolor{ForestGreen}{973}} \\
		\bottomrule
	\end{tabular}
     \vspace{-4mm}
\end{table}

\subsection{Evolution Stage Analysis}
\label{sec:results:ablation}
\noindent
Table~\ref{tab:stage-ablation} quantifies the contribution of each evolution
stage using the retained commits from the winning runs. From the default VTR-9
VPR implementation, S1 evolves packing, S2 evolves placement, S3 evolves
routing, S4 performs cross-stage optimization across multiple stages (pack,
place, route) and ports of published algorithms, and S5 tunes the exposed hyperparameters of the evolved
implementation. Different stages contribute to the CPD reduction on different
circuits, and no single stage dominates across all designs. Overall, the
code-level evolution stages S1--S4 account for 62\% of the total CPD
improvement, while S5 contributes the remaining 38\%. It's important to note here that this is tuning on the evolved code, not tuning on the stock VPR code.

\begin{table}[t]
    \centering
    \caption{Stage-wise CPD improvement of VPR-Evolve. }
    \vspace{-4mm}
    \label{tab:stage-ablation}
    \small
    \setlength{\tabcolsep}{4pt}

    \resizebox{0.85\columnwidth}{!}{%
        \begin{tabular}{l rr rrrrr}
            \toprule
            & \multicolumn{2}{c}{CPD (ns)}
            & \multicolumn{5}{c}{$\Delta$CPD by stage (ns)} \\
            \cmidrule(lr){2-3}
            \cmidrule(lr){4-8}
            Circuit
            & VTR 9
            & VPR-Evolve
            & S1
            & S2
            & S3
            & S4
            & S5 \\
            \midrule
            LU8
            & 73.40
            & 72.60
            &
            &
            &
            &
            & $-$0.80 \\

            bgm
            & 20.04
            & 19.87
            & $+$0.27
            & $-$0.29
            &
            &
            & $-$0.14 \\

            sv2
            & 13.06
            & 12.77
            &
            & $-$0.11
            &
            & $-$0.17
            & \\

            LU32
            & 75.81
            & 72.60
            & $-$1.99
            & $-$0.05
            &
            &
            & $-$1.16 \\

            mcml
            & 47.91
            & 46.84
            & $-$1.27
            & $+$0.19
            &
            &
            & \\

            \midrule
            \textbf{Total}
            &
            &
            & $-$2.99
            & $-$0.26
            & $-$0.00
            & $-$0.17
            & $-$2.10 \\
            \bottomrule
        \end{tabular}%
    }
    \par\vspace{2pt}
\parbox{\columnwidth}{\raggedright
CPD values are 5-seed
    in-loop means from each circuit's winning flow, so they differ from the
    100-seed final evaluation in Table~\ref{tab:qor}. Stage deltas sum the
    kept commits' CPD changes; positive entries are CPD regressions retained
    because they improved the composite score.}
    \vspace{-6mm}
\end{table}

\section{Discussion}
\label{sec:discussion}

\begin{table}[t]
    \centering
    \caption{Evolution artifacts: (a) a partial \texttt{mcml} evolution
    trace,  with CPD and score over the five in-loop seeds;  and (b) final
    lines of code change over default VTR-9.}
    \label{tab:evolution-artifacts}
    \vspace{-10pt}

    \begin{minipage}[t]{0.68\columnwidth}
        \vspace{0pt}
        \centering
        \textbf{(a) \texttt{mcml} evolution trace}\\
        \setlength{\tabcolsep}{2pt}
        \resizebox{\linewidth}{!}{%
            \begin{tabular}{l l c r r}
                \toprule
                Commit & Stage & Decision & CPD (ns) & Score \\
                \midrule
                \texttt{87beff686} & pack
                    & reverted & 46.65 & 999.03 \\
                \texttt{4811c323f} & place
                    & \textbf{KEPT} & \textbf{46.84} & \textbf{998.93} \\
                \texttt{16492548f} & place
                    & reverted & 47.52 & 1000.73 \\
                \texttt{1c4bbecdf} & route
                    & reverted & 46.84 & 999.73 \\
                \texttt{a1dd0eb3e} & cross-stage
                    & reverted & 47.36 & 1000.80 \\
                \texttt{8f4c9bf2c} & cross-stage
                    & reverted & 48.20 & 1002.18 \\
                \texttt{0b668797e} & tune
                    & reverted & 47.67 & 1000.77 \\
                \texttt{1acac63e4} & tune
                    & reverted & 48.47 & 1003.16 \\
                \texttt{de6b925e3} & tune
                    & reverted & 48.21 & 1003.02 \\
                \texttt{26b8b502e} & tune
                    & reverted & 50.30 & 1004.95 \\
                \texttt{3c47deeee} & tune
                    & reverted & 46.83 & 1001.49 \\
                \midrule
                \multicolumn{3}{l}{Default VTR-9 baseline}
                    & 47.91 & 1000.00 \\
                \bottomrule
            \end{tabular}%
        }
    \end{minipage}%
    \hspace{0.01\columnwidth}%
    \begin{minipage}[t]{0.30\columnwidth}
        \vspace{0pt}
        \centering
        \small
        \textbf{(b) Code changes}\\
        \setlength{\tabcolsep}{4pt}
        \renewcommand{\arraystretch}{1.05}
        \begin{tabular}{l c}
            \toprule
            Circuit & \shortstack{Lines\\changed} \\
            \midrule
            LU8  & \shortstack{$+640$\\$-12$} \\
            \cmidrule(lr){1-2}
            bgm  & \shortstack{$+352$\\$-8$} \\
            \cmidrule(lr){1-2}
            sv2  & \shortstack{$+403$\\$-34$} \\
            \cmidrule(lr){1-2}
            LU32 & \shortstack{$+172$\\$-2$} \\
            \cmidrule(lr){1-2}
            mcml & \shortstack{$+27$\\$-9$} \\
            \bottomrule
        \end{tabular}
    \end{minipage}
    \parbox{\columnwidth}{\raggedright
Bold marks the
    final state, which has the best (lowest) score.}
    \vspace{-15pt}
\end{table}

\begin{figure}[t]
	\centering
	\begin{lstlisting}[
  basicstyle=\ttfamily\tiny,
  columns=fullflexible,
  keepspaces=true,
  frame=single,
  framesep=3pt,
  xleftmargin=2pt,
  aboveskip=0pt,
  belowskip=0pt,
  moredelim={[l][\color{ForestGreen}]{+}},
  moredelim={[l][\color{red}]{-}},
]
// vpr/src/pack/greedy_candidate_selector.cpp
   float dist = get_manhattan_distance_to_tile(target_loc,
                    cluster_tile_loc, g_vpr_ctx.device().grid);
-  float gain_mult = 1.0f;
-  if (dist < appack_options.dist_th) {
-    gain_mult = 1.0f - (appack_options.quad_fac_sqr * dist * dist);
-  } else {
-    gain_mult = 1.0f / std::sqrt(dist - appack_options.sqrt_offset);
-  }
+  // [s1-appackgain] Smooth Gaussian decay replaces the piecewise
+  // quad/sqrt knee.
+  const float sigma = appack_options.gauss_sigma;  // = 2.0f
+  float gain_mult = std::exp(-(dist * dist) / (2.0f * sigma * sigma));
   VTR_ASSERT_SAFE(gain_mult >= 0.0f && gain_mult <= 1.0f);
\end{lstlisting}
\vspace{-4mm}
	\caption{Commit \texttt{87beff686}: the addition of Gaussian gain decay that had the
		largest single-commit CPD win (\texttt{mcml}, 47.91~$\rightarrow$~46.65\,ns; a later commit brings the CPD to 46.84\,ns).}
        \vspace{-6mm}
	\label{fig:diff}
\end{figure}

% \begin{wraptable}{r}{80pt}
% 	\vspace{-25pt}
% 	\centering
% 	\caption{Final lines of code changes per circuit compared to VPR.
% 		\blueHL{*** is a placeholder for now.}}
% 	\label{tab:loc}
% 	\small
% 	\vspace{-10pt}
% 	\begin{tabular}{l r}
% 		\toprule
% 		Circuit & Lines $+$/$-$    \\
% 		\midrule
% 		LU8     & $+$640\,/\,$-$12 \\
% 		bgm     & $+$352\,/\,$-$8  \\
% 		sv2     & $+$403\,/\,$-$34 \\
% 		LU32    & $+$172\,/\,$-$2  \\
% 		mcml    & $+$27\,/\,$-$9   \\
% 		\bottomrule
% 	\end{tabular}
% \end{wraptable}

\noindent
VPR-Evolve produces not only a final evolved algorithm, but also an annotated
trace of the evolution process. Table~\ref{tab:evolution-artifacts}(a) shows a partial
trace for \texttt{mcml}, including the stage of each attempted modification,
whether it was retained, and its measured CPD and composite score.
This trace makes the search auditable: retained changes show how the final
implementation was constructed, while reverted changes reveal directions that
were evaluated but not used. 

Fig~\ref{fig:diff} shows part of the diff between the evolved and the
original VPR source that produced the single largest CPD improvement found for
\texttt{mcml}. The evolved packing code replaces VPR's piecewise distance
attenuation with a smooth Gaussian decay, reducing the five-seed in-loop
CPD from 47.91\,ns to 46.65\,ns. The score gate then retained one further
placement change that gives back 0.19\,ns of CPD but improves the composite
score, so the final implementation lands at 46.84\,ns, a 2.24\% net CPD
reduction. The Gaussian-decay edit is an algorithmic
change rather than a new setting of an existing parameter. The retained edit
adds six lines and removes six lines across two files. Table~\ref{tab:evolution-artifacts}(b) summarizes the final source-code differences for each
circuit. The patch size varies from only \(+27/-9\) lines for \texttt{mcml} to
\(+640/-12\) lines for \texttt{LU8}, and does not increase with circuit size.
Indeed, the largest circuit produces the smallest patch and the best composite
score. These results show that design-specific algorithm evolution need not
produce an entirely new tool; it can identify localized, reviewable changes to
an existing implementation.

\section{Conclusion}
\label{sec:conclusion}
\noindent
We presented VPR-Evolve, a multi-agent framework that specializes VPR to a
target FPGA design by evolving its packing, placement, and routing source code.
Planner, Coder, Reviewer, and Inspiration Collector agents cooperate through a
shared memory, and every candidate is judged by a full VPR build and run against
a composite objective over CPD, WL, and RT. Across five VTR~9 benchmarks, VPR-Evolve improves the
composite score by up to 2.7\% over stock VPR, reducing CPD by up to 9.8\%, WL
by up to 18.1\%, and RT by up to 79.3\%. Compared with an
AutoTuner hyperparameter-tuning baseline given a larger evaluation budget, it
reduces CPD by up to 6.0\%, WL by up to 2.2\%, and RT by up to 7.8\%.
Code-level evolution stages account for most of the CPD gain, while the
resulting patches remain small and reviewable, and many transfer to other
designs. These results show that design-adaptive algorithm evolution can move
beyond the QoR frontier of a fixed CAD algorithm and toward design-tool
co-exploration for FPGA physical design.

\bibliography{bib/bibfile}

\end{document}